\documentclass[aps,prc,showpacs,twocolumn,floatfix,nofootinbib]{revtex4}
\usepackage{epsfig,graphics,color}
\usepackage{graphicx}
\usepackage{dcolumn}
\usepackage{bm}
\usepackage{amsmath}

\newcommand \bvec{\left( \begin{array}{c} }
\newcommand \evec{\end{array} \right)}

\newcommand \bea{\begin{eqnarray} }
\newcommand \eea{\end{eqnarray} } 
\newcommand \nn {\nonumber}
\newcommand {\be} {\begin{equation}}
\newcommand {\ee} {\end{equation}}

\begin{document}

\title{Jets in $d(p)$-$A$ Collisions: Color Transparency or Energy Conservation }

\author{Michael Kordell II}
\author{Abhijit Majumder}
\affiliation{Department of Physics and Astronomy, Wayne State University, Detroit, MI 48201.}

\date{\today}

\begin{abstract} 
The production of jets, and high momentum hadrons from jets, produced in deuteron ($d$)-$Au$ collisions at the relativistic heavy-ion collider (RHIC) and proton ($p$)-$Pb$ collisions at the large hadron collider (LHC) are studied as a function of \emph{centrality}, a measure of the impact parameter of the collision. 
A modified version of the event generator PYTHIA, widely used to simulate $p$-$p$ collisions, is used in conjunction with a nuclear Monte-Carlo event generator which simulates the locations of the nucleons within a large nucleus. 
We demonstrate how events with a hard jet may be simulated, in such a way that the 
parton distribution function of the projectile is frozen during its interaction with the extended nucleus. Using our approach, we demonstrate that the puzzling 
enhancement seen in peripheral events at RHIC and the LHC, as well as the suppression seen in central events at the LHC are mainly due to \emph{mis}-binning of central and semi-central events, containing a jet, as peripheral events. This occurs due to the suppression of soft particle production away from the jet, caused by the depletion of energy available in a nucleon of the deuteron (in $d$-$Au$ at RHIC) or in the proton (in $p$-$Pb$ at LHC), after the production of a hard jet.  We conclude that partonic correlations built out of simple energy conservation are mostly responsible for such an effect.
\end{abstract}

\pacs{25.75.-q,13.87.-a,25.45.-z}

\maketitle

\section{Introduction}

In the context of hard processes in heavy-ion collisions, maximally asymmetric collisions, such as $d$-$Au$ at RHIC and $p$-$Pb$ at the LHC,  
have served the purpose of baseline measurements: Quantifying initial state nuclear effects without the presence of a hot-dense extended final state. 
Early measurements of suppressed back-to-back hadron correlations, with momenta perpendicular to the colliding nuclei, at the STAR detector at RHIC~\cite{Adler:2002tq} for $Au$-$Au$ collisions, compared with a null effect in 
$d$-$Au$ (compared with $p$-$p$) established jet quenching as a final state effect that takes place 
primarily in the presence of an extended Quark-Gluon-Plasma (QGP). These jets with momentum transverse to the incoming beams, were quenched in $Au$-$Au$, but were minimally affected in $d$-$Au$ collisions.

These were consistent with measurements of a lack of suppression in the expected yield of high transverse momentum (leading) hadrons in 
$d$-$Au$ collisions at the PHENIX detector~\cite{Adler:2003ii}. In 2006, the PHENIX collaboration, extended this analysis to centrality (the experimental measure of impact parameter) dependent suppression~\cite{Adler:2006bv,Adler:2006wg}. This data demonstrated an odd enhancement in the yield of high momentum hadrons in peripheral 
$d$-$Au$ events. While nuclear effects which modify the dynamics of jet production, were expected in central events, where nucleons from the deuteron encounter several collisions with the large nucleus, these were not expected at all in peripheral events where the deuteron has fewer collisions with the large nucleus.

Recently there have been a series of new measurements, both by PHENIX~\cite{Adler:2006wg} at RHIC 
and by ATLAS  at the Large Hadron Collider (LHC), on the spectrum of 
high transverse momentum (high $p_{T}$) jets produced in $d$-$Au$~\cite{Wysocki:2013caa} and $p$-$Pb$~\cite{Perepelitsa:1748487} collisions. 
The measurements plot the centrality dependent nuclear modification factor $R_{dAu}$ of high $p_T$ jets: 
A ratio of the detected yield of jets to that expected based on an estimate of the number of nucleon-nucleon collisions in one $p(d)$-$A$ collision. 
In both cases, one notices an enhancement in the $R_{d Au}$ in peripheral events and a ``suppression''  in central collisions. 
In a study of the rapidity dependence of the reconstructed jet, by the ATLAS collaboration, it was observed that this peripheral enhancement and central suppression was 
much more prevalent in the $p$ going direction and vanishing in the $Pb$ direction. 

These results are rather counterintuitive. Nuclear effects, in particular those that involve jets and jet production, are expected to be dominant in central events where the initial state engenders several nucleon-nucleon collisions, also 
the final out going partons have to traverse a more extended medium. Similar arguments may be ascribed to the rapidity dependence of hard particle production, with hard partons 
traversing longer distances in the nucleus going direction than in the $p$ or $d$ going direction. 

In this paper we posit that events which lead to the production of a hard jet, requiring an initial state parton with a considerable value of $x$, have initial states with a fewer number of soft partons, due to the large amount of energy that has been drawn away from the nucleon by the high-$x$ parton. 
This effect is most pronounced on the partons in the $p(d)$ going direction, and much less on the $A$ going direction as the formation of a hard parton in a single nucleon (in a nucleus) does not effect the soft parton distribution in the remaining nucleons. 
The higher the $x$ required, the more the suppression in the soft particle production. 
Thus reactions with very high energy jet production probe the correlation between partons within a nucleon. This sensitivity to multi-parton hard-soft correlations is unique to these experiments, which probe a hitherto unmeasured facet of nucleon structure: is there a strong correlation between the $x$ values of the leading partons in a given event and the total number of partons in the nucleon, in that event. By ``strong'' we are suggesting something more than the trivial correlation due to straightforward energy conservation: is there a kind of \emph{color transparency} in the initial state, for events with a hard jet in the final state? Our calculations do not provide a clear answer to this second question.
Beyond this, another goal of this work is to provide a reliable parameter free event generator which may be used, with certain caveats, to reproduce at least some portion of these 
new data on $p(d)$-$A$ collisions with jet production. The results of this paper will provide detailed input to a more dedicated event generator that will have to be constructed to study such collisions in greater detail. 

In the remainder of this paper, we describe our model and how soft particle production is affected by the production of a hard jet. To make direct connection to experiments, we set out to modify the PYTHIA event generator~\cite{Sjostrand:2014zea} which is used extensively to model $p$-$p$ collisions. 
To date there have been several approaches which have attempted to describe this new striking physics result. In Ref.~\cite{Bzdak:2014rca}, the authors have proposed a similar mechanism of enhancement in peripheral events and suppression in central events but not incorporated it in an event generator framework.
In Ref.~\cite{Alvioli:2014eda}, the authors have proposed that the wave-function of the proton is considerably modified in the presence of a hard parton. In Ref.~\cite{Armesto:2015kwa}, the authors have attempted to understand the effect of the energy depletion due to jet formation using the HIJING event generator~\cite{Wang:1991hta,Gyulassy:1994ew}. In none of these calculations, could the authors achieve widespread agreement with the data.  
The current effort has been constructed entirely within the PYTHIA event generator, by modifying it. As such, we suffer from several constraints which are inbuilt within this particular event generator. The reader may question why we did not use the HIJING event generator as in Ref.~\cite{Armesto:2015kwa}, the primary reason behind this is the resampling of the parton distribution function between collisions; this has the effect of the proton 
(or nucleon in $d$-$Au$ collisions) changing its parton distribution function between successive collisions which changes the distribution of soft partons that arise after the hard parton has been extracted. 

In the subsequent section we describe the event generator that samples the location of the nucleons in the two incoming nuclei. In Sec.~\ref{pdf}, we outline the changes introduced into the PYTHIA event generator. In Sec.~\ref{expt} we present comparisons with experimental data at RHIC and LHC. Our conclusions are presented in Sec.~\ref{outlook}.

\section{Sampling the nuclear distribution }\label{nuclear}

Maximally asymmetric collisions such as $p$-$Pb$ or $d$-$Au$ represent cases where the experimentally determined centrality of the event appears to be influenced by the production of a hard jet. In order to simulate jet production in such systems, the PYTHIA event generator was modified and extensively 
used. This modification of the event generator depended on the number of nucleon-nucleon collisions in a given $p(d)$-$A$ event. This number of collisions was determined using several methods. In this section we describe these methods. 

Along with a description of our setup, we will explore and 
eliminate the most na\"{i}ve explanation of the observed correlation between jet production and centrality: The deuteron due to its large size, often has the 
proton and neutron far apart and thus cases where a jet is most likely to be produced, when either nucleon strikes the densest part of the oncoming nucleus may coincide with cases where the other nucleon simply escapes without interaction leading to reduced soft particle production. 
It should be pointed out, in passing, 
that such a scenario is immediately ruled out by an almost identical correlation between jet production and centrality in LHC collisions, 
where there is only one proton colliding with the  large nucleus.

\subsection{The Deuteron:}

Collisions at the LHC always involve a proton colliding with a $Pb$ nucleus. 
However, at top RHIC energies the collisions are usually that of a deuteron $(d)$ on a $Au$ nucleus. 
The deuteron is an extremely well studied state in low energy nuclear physics. 
The wave-function of the deuteron is given by the Hulth\'{e}n form~\cite{greiner1996nuclear}: 
\bea
\psi_{H}(r) = \frac{e^{-ar} - e^{-br}}{r}, 
\eea
where, $a=0.228$/fm, $b=1.18$/fm. 
The probability distribution of a nucleon within a deuteron is given as, 
\bea
\rho(r) = | \psi_{H} (r) |^{2}.
\eea
This distribution is sampled to obtain the positions of the two nucleons. 

\begin{figure}[htbp]
\resizebox{3in}{2.5in}{\includegraphics{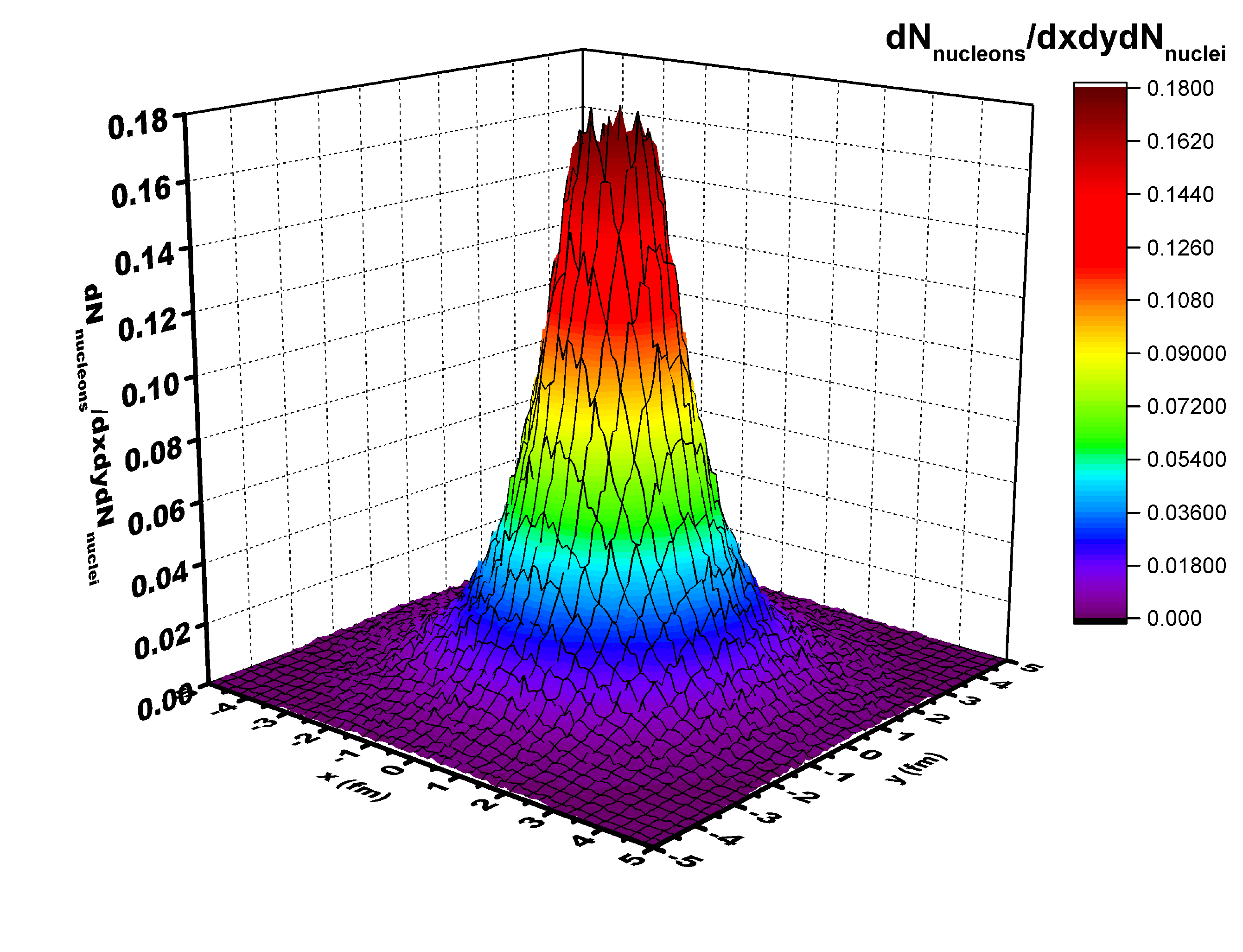}}
    \caption{Color Online: The sampled Hulth\'{e}n distribution for two nucleons in a deuteron. }
    \label{fig1}
\end{figure}

As is well known, the Hulth\'{e}n wave-function leads to a rather wide nuclear distribution. 
This is illustrated in Fig.~\ref{fig1}, where we plot three representative events, with both the $Au$ nucleus and 
the deuteron distributions projected on the $z$-axis which is the axis of momentum of the two nuclei. 
As can be seen 
from Fig.~\ref{fig1}. the nucleons in the deuteron may be close together, as well as, a gold radius 
apart. Due to the large separation between the nucleons, excluded volume corrections were unnecessary but were still included.

\begin{figure}[htbp]
\resizebox{3in}{2.5in}{\includegraphics{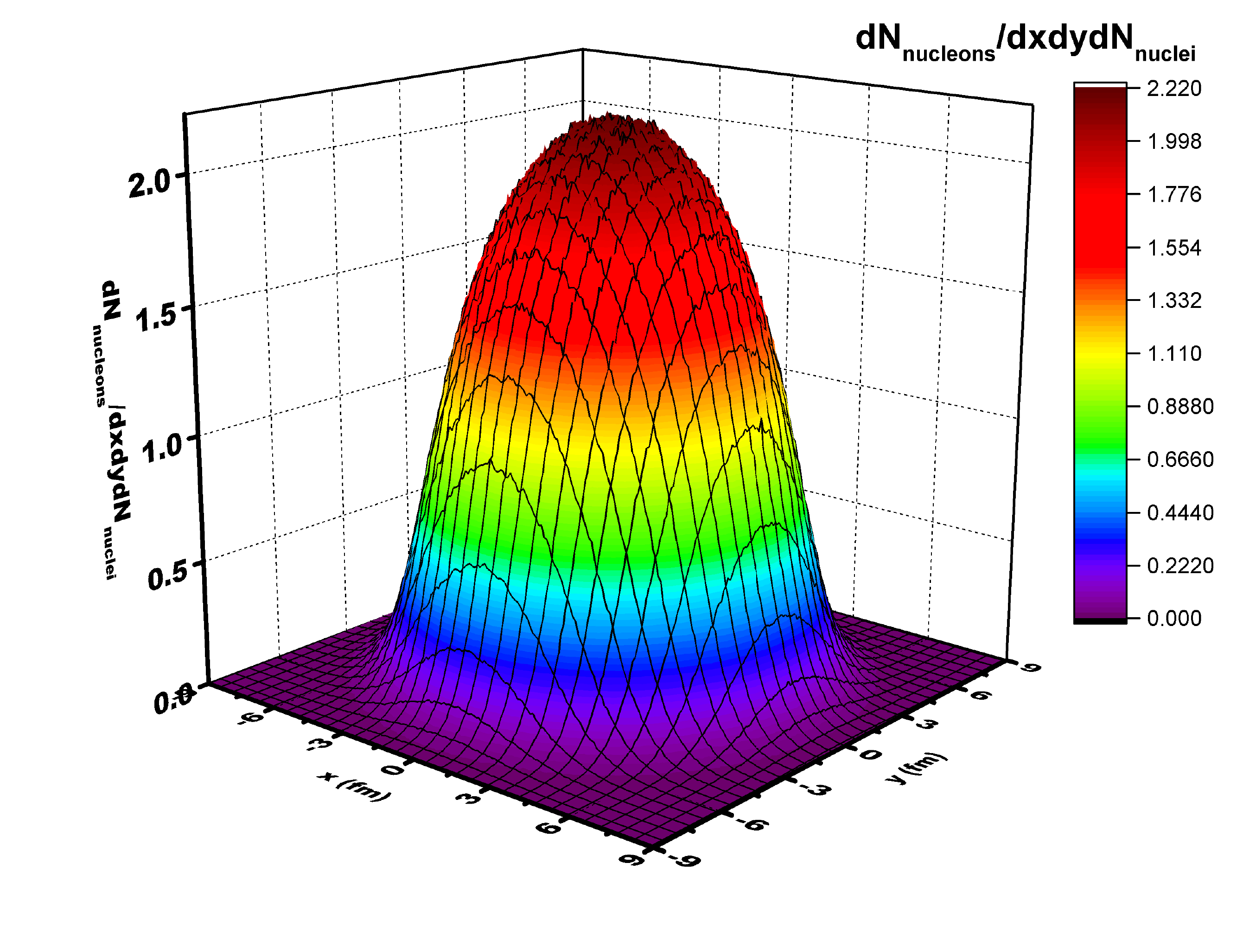}}
    \caption{Color Online: The sampled Woods-Saxon distribution for a large nucleus (in this case $Au$ with an $A=192$.)}
    \label{fig2}
\end{figure}

\begin{figure}[htbp]
\resizebox{3in}{4in}{\includegraphics{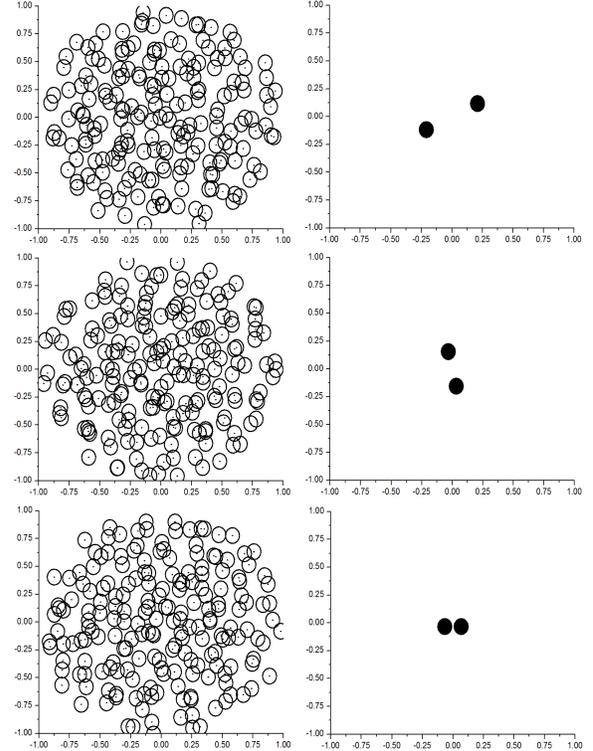}}
    \caption{Three separate events in $d$-$Au$ collisions. Nucleon distributions are projected onto the $x$-$y$ plane.}
    \label{fig3}
\end{figure}

\subsection{The Large Nucleus ($Au$ or $Pb$)}

Moving to the nuclear state, there are several methods that may be used to simulate the fluctuating initial state represented 
by the large nucleus. 
In this work, we will only focus on the $Au$ or $Pb$ nucleus, as these are studied experimentally. In most cases, 
we will use the Woods-Saxon density distribution, given at a radial distance $r$ as:
\bea
\rho(r) = \frac{\rho_0}{1 + e^{(r-R)/a}}, \label{WSdistribution}
\eea
 where $\rho_0$ is a constant related to the density at the center of the nucleus, $R$ is the radius of the nucleus, and $a$ is the 
 skin depth. These parameters are chosen to match those used by the experiments at RHIC and LHC 
 (for $Au$, $a = 0.535$fm, $R = 6.38$fm; for $Pb$, $a = 0.546$fm, $R = 6.62$fm). 
 The Woods-Saxon distribution of Eq.~\eqref{WSdistribution} is a single particle distribution. On top of this we introduce a nucleon-nucleon 
 correlation by hand: the excluded volume correction. This is done similar to the method of Ref.~\cite{Majumder:1998qz}, where we generate a 
 set of 3 random numbers which isolate the location of a nucleon. If this location is within an exclusion distance of $d=2 R_p$ (twice the proton radius) 
 of another nucleon, then this location is abandoned and another generated. The process is continued until all $A$ nucleons have been included. 
 At the end of this process the center-of-mass of the nucleus is calculated and the nucleus is re-centered.
 
While only the Hulth\'{e}n form is used for the deuteron, several probability distributions beyond Woods-Saxon were tried for the nucleon distribution in a large nucleus. These include distributions based on shell-model wave-functions both with and without a modified delta interaction to account for the short range repulsion between nucleons in a nucleus~\cite{brussaard1977shell} (simple excluded volume).  However, none of these enhancements led to any noticeable changes in the final results as compared to the Woods-Saxon distribution with a simple excluded volume. It should be pointed out that in this effort, we have only considered $p$-$A$ and $d$-$A$ collisions which only sample the single and two-nucleon distribution within a nucleus. It is entirely possible that the collision of nuclei larger than a deuterium with nuclei smaller than $Au$ may lead to the greater role for multi-particle correlations within a nucleus. There is very little information in nuclear structure on such multi particle correlations. We will not discuss this issue further in this effort, and only focus on simulations using the Woods-Saxon distribution with an excluded volume.

\subsection{Transverse Size of Nucleons and Binary Collisions:}

The nuclear Monte-Carlo generator samples nucleons from the $Au$ (or $Pb$) side and from the $d$ side and then projects these on the 
$x$-$y$ plane as shown in Fig.~\ref{fig3}. In the work presented in this paper, the transverse size of the nucleons has not been modified with 
the energy of the collision. The inelastic cross section for nucleon-nucleon scattering is known to grow with collision energy. While centrality selection at 
the nuclear level is one of the major issues dealt with in this effort, no centrality selection is imposed on the individual nucleon-nucleon encounters. 
As a result, when a proton from the $d$ overlaps with another
from the $Au$ side, no matter how small the overlap, the entire parton distribution function (PDF) of either nucleon is enacted in the collision, i.e., nucleon-nucleon collisions are not expected to 
have any centrality dependence. 

Glancing at Fig.~\ref{fig3}, it becomes clear that if the transverse size of the nucleons is increased with increasing energy then this will lead to an increase in 
the number of binary collisions and that will lead to an artificial excess enhancement of the particle 
production from each individual nucleon-nucleon collision. In this work, we will use the event generator PYTHIA to simulate nucleon-nucleon collision. 
Within the PYTHIA event generator the cross section increases with energy. 
To counter the possible artificial increase in particle production with energy, the full cross section generated by PYTHIA is used with no change in the 
geometric size of the nucleon with the energy of the nuclear collision. In a future effort, an impact parameter in nucleon-nucleon collisions will be used to generate particle 
production in events where the two nucleons do not overlap completely. 

Once both nuclei have been generated, and centers of mass determined, the impact parameter $b$ is simulated with a probability distribution $dP/db^2 = 1/b_{Max}^2$, and 
the angle of the impact parameter is determined randomly between $0$ and $2 \pi$. 
The maximal impact parameter $b_{Max}$ is chosen such that no dependence is observed in minor changes of this quantity. 
There is no further reorienting of the nuclei. The number of binary collisions 
can now be determined by simply counting the number of nucleons in the $Au$ side, whose centers are within a transverse distance $d = 2 R_p$ of a nucleon in the deuteron.
There arise events where not a single collision takes place, these events are dropped from the analysis.

\begin{figure}[htbp]
\resizebox{3.5in}{3in}{\includegraphics{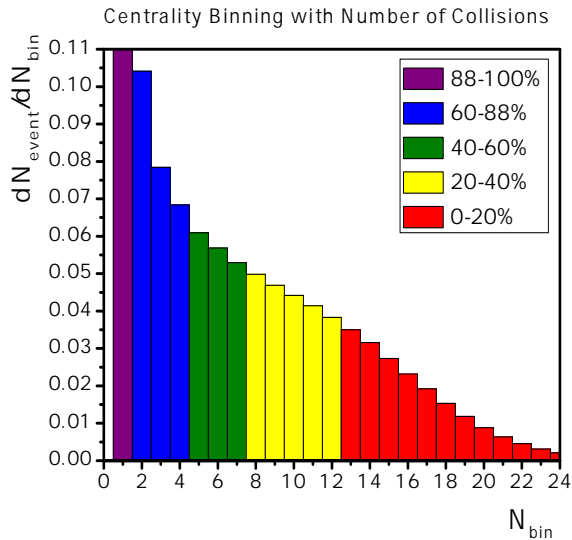}}
    \caption{Color Online: The event distribution in $d$-$Au$ collisions as a function of the number of binary collisions and the division of events in the four different centrality bins.}
    \label{fig4}
\end{figure}

Based on the above considerations, 
we present the results of the nuclear Monte-Carlo simulations for a $d$-$Au$ collisions in Fig.~\ref{fig4}.
In Fig.~\ref{fig4}, the distribution of events as a function of the number of binary collisions is presented. 
Following this, events are divided into 4 bins (0-20\%, 20-40\%, 40-60\%, 60-88\%) based on the fraction of the total number of events 
contained in these bins. Each of these bins in the number of collisions represents a range of overlapping impact parameters. 
While this represents the standard method of determining centrality in theoretical calculations or simulations, 
we will show, in a later section, that this method of determining the centrality of the event leads to results that are not consistent with 
experimental results for high transverse momentum (high-$p_T$) pion, charged particle, and jet production at both RHIC and LHC energies. 

In this section we have focussed mostly on $d$-$Au$ collisions where both incoming nuclei have to be simulated. In subsequent sections we will 
also present results for $p$-$Pb$ collisions where only one nucleus needs to be simulated. There are no other considerations concerning $p$-$Pb$ 
that need to be made other than the location of the $p$ is set by the impact parameter $b$. As pointed out above, no explicit change in the transverse 
size with energy has been used in this first attempt to understand the behavior of jets in $p(d)$-$A$ collisions. We also mention that in all simulations, 
we keep track of the isospin of the nucleons.


\section{The Modified Parton Distribution Function} \label{pdf}

Using the nuclear collision event generator, the number of nucleon-nucleon collisions in each event may be determined. 
Each nucleon in the deuteron, in a $d$-$Au$ collision at RHIC, or the proton in $p$-$Pb$ collisions at the LHC, will potentially 
engender several collisions with nucleons in the large nucleus. At RHIC the relativistic $\gamma$ factor is about a 100 while it is 
close to $2750$ at the LHC. At such large boosts, the parton distribution function within the nucleon is time dilated to distances well 
beyond the length of the large nucleus. As a result, the parton distribution of the nucleons in deuteron (in a $d$-$Au$) collision, or that in 
the proton in $p$-$Pb$ collisions is ``static'' (frozen) as it progresses through the large nucleus. We use the word static to indicate that the 
parton distribution, though being continuously depleted by collisions with partons in the nucleons from the large nucleus, is itself 
not undergoing any intrinsic fluctuation in the course of its passage through the large nucleus. 

This brings the discussion to the primary point of this paper: Consider the case, where, in the course of fluctuations of the PDF, the proton in 
$p$-$Pb$ (or one of the nucleons in $d$-$Au$) has focussed a large amount of energy within a single parton. This parton, in collision with a 
similar parton in the oncoming nucleus will produce back to back jets at mid-rapidity. The presence of a parton with such a large energy will 
lead to less energy being available for the production  of other softer partons. As a result, there will be a depletion in the number of soft partons
in the proton in $p$-$Pb$ (or projectile nucleon in $d$-$Au$) collisions. A similar situation will occur in one of the nucleons within the large nucleus.
As a result, an event with a jet will lead to the production of fewer charged particles. 

To simulate this effect, we treat the collision of the $p$ (or any of the nucleons in $d$) with a string of $n$ nucleons in the large nucleus as a single collision between a
nucleon and an object with a larger (modified) PDF. As a result, the PDF of the projectile nucleon is sampled only once. To be clear, there are several methods to carry this 
out, we will only focus on the most expeditious method. In the remainder of this paper, we will refer to the collection of $n$ nucleons struck as 
a single entity by the projectile nucleon as a ``super''-nucleon.

\begin{figure}[htbp]
\resizebox{3.5in}{3in}{\includegraphics{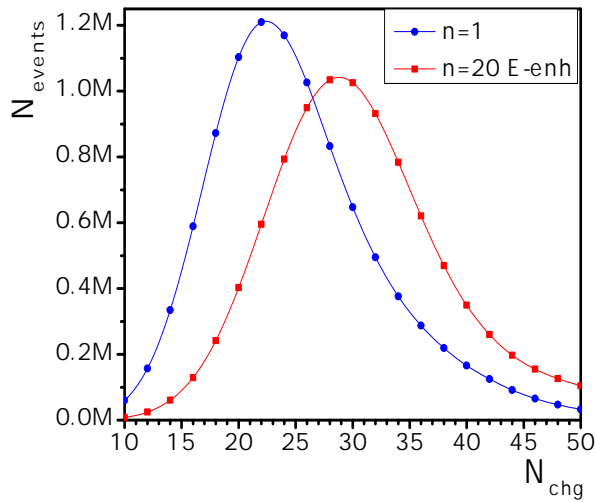}}
    \caption{Color Online: The multiplicity of charged particles in a simulated $d$-$Au$ collision with the $Au$ side simulated as 
    a super-nucleon with a parton distribution function given as $F_S (x) = n_p F_p(x) + n_n F_n(x)$,  and energy enhanced as $E_S =  (n_p + n_n) E$.
    In the above plot $n_p = 10$, and $n_n = 10$.}
    \label{fig5}
\end{figure}

As a first step to simulate the super-nucleon, we enhance the PDF of one of the incoming nucleons as $F_S (x) = n_p F_p(x) + n_n F_n(x)$. Where, $n_p$ and $n_n$ 
are the number of protons and neutrons struck by the projectile nucleon. 
Along with this the energy 
of the super-nucleon is also enhanced as $E_S =  (n_p + n_n) E$, where $E$ is the energy of the projectile nucleon in the lab frame. This prescription turns out to 
produce a very faithful description of the soft particle production in $d$-$Au$ (or $p$-$Pb$) collisions. This is illustrated by the increase in the yield of soft particles 
with increasing enhancement of the super-proton shown in Fig.~\ref{fig5}. One notes both an increase in the mean value of charged particle production, as well as an 
increase in the event-by-event fluctuation in charged particle production, as expected. 

\begin{figure}[htbp]
\resizebox{3.5in}{3in}{\includegraphics{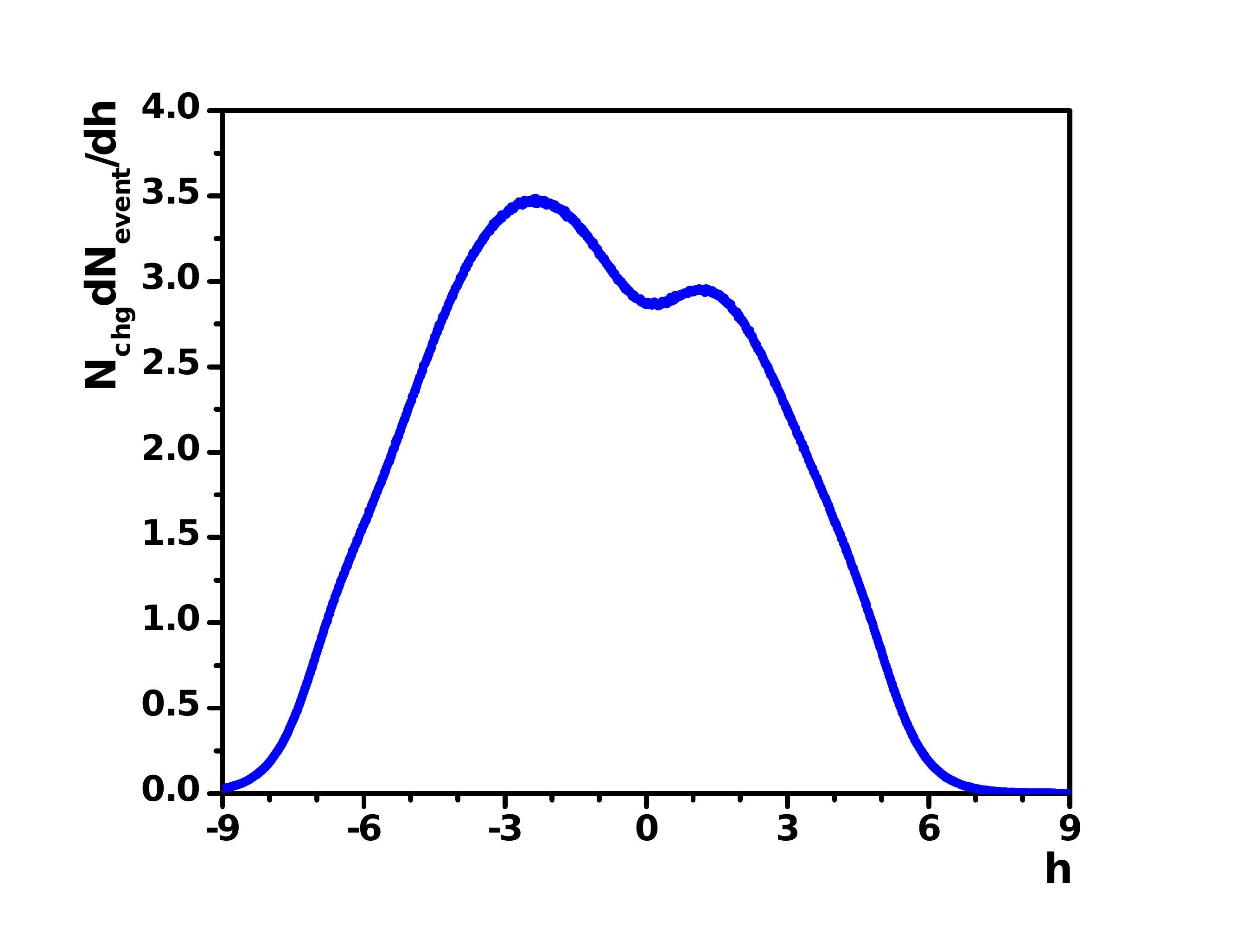}}
    \caption{Color Online: The pseudo-rapidity distribution of charged particles in a simulated $d$-$Au$ collision with the $Au$ side simulated as 
    a super-nucleon with a parton distribution function given as $F_S (x) = n_p F_p(x) + n_n F_n(x)$,  and energy enhanced as $E_S =  (n_p + n_n) E$.}
    \label{fig6}
\end{figure}

Yet another feature of this formula for the super-nucleon is that it also gives a rather faithful representation of the pseudo-rapidity distribution of the 
produced charged particles. This distribution for minimum bias events, plotted in Fig.~\ref{fig6}, shows the ``classic'' asymmetric double humped structure of the pseudo-rapidity distribution for $d$-$Au$ collisions at RHIC energies. The overall normalization is less than that measured in actual experiments. However, one should 
recall that we are generating this by modifying PYTHIA where only the interactions between the projectile nucleon and the column of struck nucleons is included. 
No re-interaction of the produced particles with the remainder of the nucleus is included, and this leads to an obvious depletion in overall particle production. 
The assumption being made in comparing these results to experimental data is that even though the overall number of charged particles (or transverse energy) 
produced is not matched between the simulations and the experiment, the relative distribution between centrality bins will be the same as in the experiment. 

In spite of the success in soft particle production using the prescription of enhancing both the PDF and the the energy of a nucleon in the target nucleus, this 
procedure leads to an uncontrollable modification to the high momentum (large-$x$) portion of the PDF. This is to be expected, as the super-nucleon now 
has $n = n_p + n_n$ times the energy of a single nucleon, and can thus produce hard partons of higher energy 
(higher even than the kinematic bound of $100$ GeV at RHIC, or $2.75$ TeV at the LHC) 
without the penalty of a rapidly falling PDF. As an illustration of this effect,  we plot the ratio of a gluon spectrum from a super-nucleon to that from a regular 
nucleon as a function of the ratio of the energy of the gluon to that of the projectile nucleon (un-enhanced nucleon). 

\begin{figure}[htbp]
\resizebox{3.5in}{3in}{\includegraphics{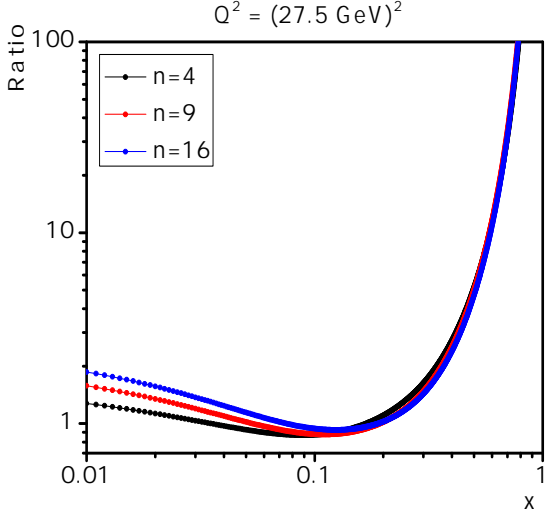}}
    \caption{Color Online: The ratio of the gluon distribution in a super-nucleon to that in a nucleon as a function of $x$,  the energy fraction of the gluon, relative to the 
    projectile nucleon.}
    \label{fig7}
\end{figure}

It is interesting to note that the soft gluon ($x<0.1$) production in enhanced in the super-nucleon as a function of the total enhancement coefficient $n = n_p + n_n$.
There is no enhancement for intermediate energy gluons $x \sim 0.1$, and then an almost $n$ independent enhancement for higher energy gluon with $x > 0.1$.
Note that this will of course be broken as one moves past the $x \geq n $, however, since the denominator of the ratio plotted in Fig.~\ref{fig7} will vanish, this 
cannot be plotted in the manner of Fig.~\ref{fig7}.

Due to this large enhancement in the hard portion of the PDF, this straightforward enhancement of the PDF for a super-nucleon cannot be used. Since the 
primary focus of the simulations reported in this paper has to do with jet production and its ensuing effect on soft particle production due to energy conservation, 
we will insist on keeping the jet production cross section as close to the reality as possible, and not enhance the energy of the super-nucleon.
For comparison with experiment, we will use a more complicated enhancement formula for the super-nucleon, where the soft portion of the PDF is modified by 
a shadowing function, and an enhancement by the number of collisions $n=n_p + n_n$, but no energy enhancement. We will also use a shadowing function which modifies the super-nucleon PDF event-by-event, depending on the number of nucleons struck by the projectile nucleon. In the case of a $d$-$Au$ collision, both nucleons may strike multiple nucleons and thus both collisions would be modeled as 
a nucleon super-nucleon collision. The formula use for this is given as, 
\bea
S(x) = 1 + ( R(x) - 1)\frac{N_{coll}}{\langle N_{coll} \rangle}, 
\eea
where $N_{coll}$ is the number of collisions encountered by a single projectile nucleon 
as it passes through the target nucleus in a given event. The mean number of collisions 
per projectile nucleon is given as $\langle N_{coll} \rangle$.
The shadowing factor of $R(x)$ which depends on $x$ and the mass number of the target nucleus $A$, is 
taken from Ref.~\cite{Li:2001xa}. For the case of a quark it has a rather involved form: 
\bea
R_q^A &=& 1 + 1.19\log^{1/6}A (x^3 - 1.2 x^2 + 0.21x) \\
&-& 0.1(A^{1/3}  - 1)^{0.6}(1 - 3.5\sqrt{x}) \exp(-x^2/0.01). \nn
\eea

\begin{figure}[htbp]
\resizebox{3.5in}{3in}{\includegraphics{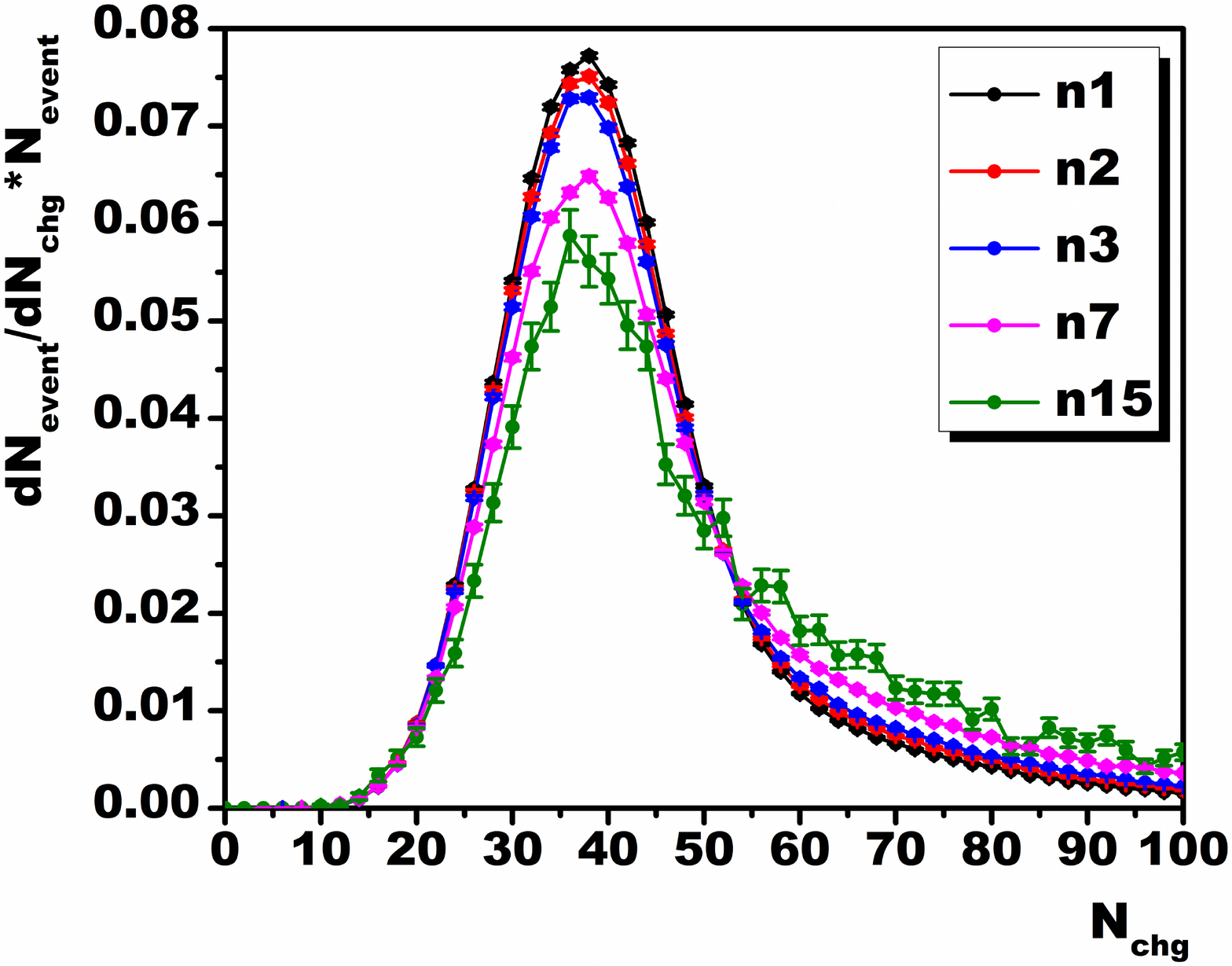}}
    \caption{Color Online: The distribution of charged particles produced in a $p$-$Pb$ collision, as a function of the 
    number of collisions suffered by the projectile proton.}
    \label{fig8}
\end{figure}

Without the enhancement in energy of the super-nucleon one does not get the asymmetric distribution of produced charged 
particles as shown in Fig.~\ref{fig6}. However, there is still an enhancement in the production of charged particles with increasing 
number of collisions. This is illustrated for $p$-$Pb$ collisions in Fig.~\ref{fig8} where the distribution of the number of charged 
particles per event is plotted for different number of collisions encountered by the proton. We present this plot for $p$-$Pb$ at 
LHC energies as the effect of our modifications to the super-nucleon PDF has a smaller effect at these energies (the shift is proportionately 
larger at RHIC).

\begin{figure}[htbp]
\resizebox{3.5in}{3in}{\includegraphics{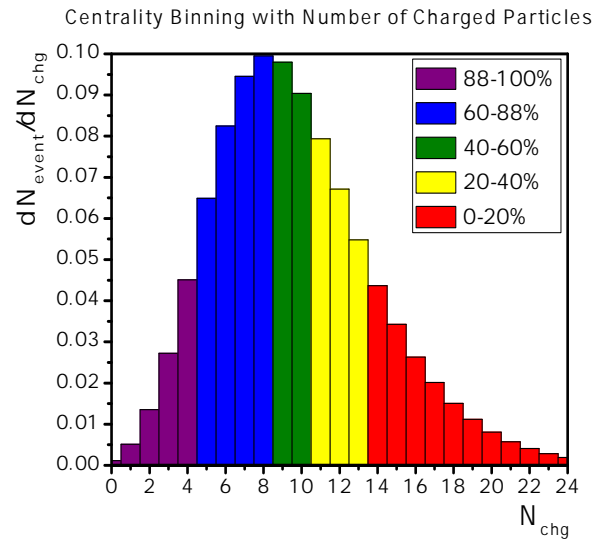}}
    \caption{Color Online: The distribution of the number of charged particles produced and the division of the events in 
    the four different centrality bins.}
    \label{fig9}
\end{figure}

No doubt, this enhancement is proportionately less than that in Fig.~\ref{fig5}, however, it is sufficient to allow us to bin in 
centrality. We point out once again, that in this process of simulating $d$-$Au$ collisions at RHIC energies, or $p$-$Pb$ collisions at 
LHC energies, the soft particle production is in no way commensurate with that in a real $d$-$Au$ or $p$-$Pb$ collision. We are 
carrying out this exercise to demonstrate the effect of a shift in centrality due to the production of a hard jet. We insist on the 
jet production cross section being unchanged in PYTHIA, while assuming that the reduced soft particle production in this model, as 
a function of the deduced centrality, is 
proportional to the particle production in a real collision. 

To illustrate this issue, we plot the distribution of the number of events as a function of the number of produced charged particles 
in our simulations for $d$-$Au$ collisions at RHIC energies. As is clearly demonstrated by this figure, there are clear, non-vanishing 
ranges of particle production, which can be clearly demarcated as centrality bins. These simulations are all done using the 
\emph{Hard-QCD} switch of PYTHIA. This is the case both for the particle production in general and for particle production in addition 
to the production of a hard jet. This is done so that the mechanisms that lead to soft particle production both in the presence and absence of 
a hard jet remain the same in the simulation. 

In what follows, we will consider jet and leading hadron production at high-$p_T$ and compare the effect of this on soft particle 
production. This will be done both for RHIC and LHC energies, for jet production at central rapidities. Charged particle detection, 
which leads to a centrality determination, will be carried out at all rapidities, i.e., over the entire collision. In actual experiments, 
charged particles are detected at rapidities far from where the jets are produced, in an effort to remove any correlation between the 
two processes. Since, in our simulated collisions, the number of particles produced is far fewer than an actual experiment, we collect 
charged particles at all rapidities, to allow to distinguish between different centralities with higher statistics.

\section{Comparison with Experiment} \label{expt}

In the proceeding sections, the model used to simulate jet (and high $p_T$ particle) production as a function of centrality in $d$-$Au$ 
collisions at RHIC and $p$-$Pb$ collisions at the LHC, was described in detail. As stated before, our primary goal is two fold: to set up 
an event generator that may be used to faithfully represent the experimental data on hard soft correlations in asymmetric collisions, 
albeit with some caveats, as well as to understand the underlying cause of the starting results in such correlations using this new event generator. 

\begin{figure}[htbp]
\resizebox{3.1in}{2.6in}{\includegraphics{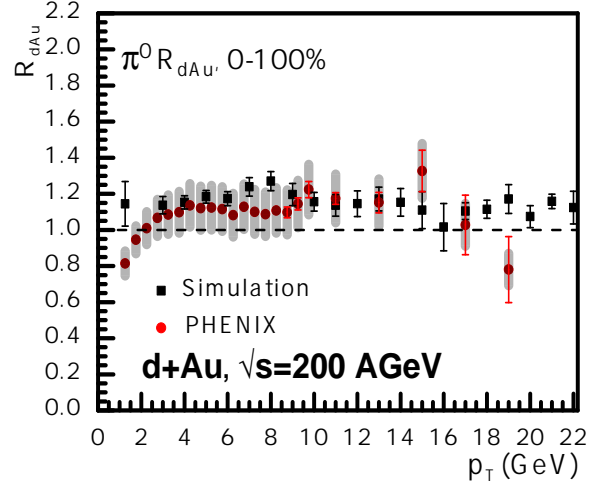}}
    \caption{Color Online: The nuclear modification factor for neutral pions for minimum bias $d$-$Au$ collisions at RHIC. Experimental 
    data are taken from Ref.~\cite{Sahlmueller:2012ru}}
    \label{fig10}
\end{figure}

Viewed in the lab or center-of-mass frame, it became clear that the nucleon PDF from both the projectile and the target are time dilated, and 
as such, cannot fluctuate in the short duration of the collision. This led us to abandon HIJING~\cite{Wang:1991hta,Gyulassy:1994ew}, and 
design a new event generator by modifying the PDF of one of the nucleons in a PYTHIA nucleon-nucleon collision. The effects of 
different modifications within PYTHIA and the overarching nuclear event generator were highlighted in the preceding sections. 
In the following, we demonstrate the successes and shortcomings of this new event generator when compared with actual experimental 
data. 

\begin{figure}[htbp]
\resizebox{3.1in}{2.6in}{\includegraphics{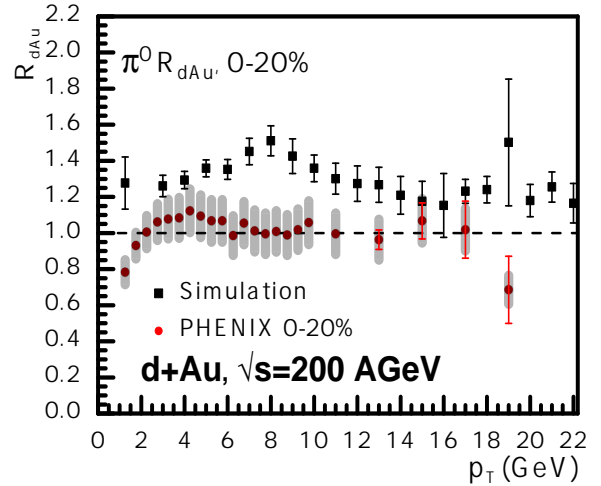}}
    \caption{Color Online: The nuclear modification factor for neutral pions for 0 - 20\% most central $d$-$Au$ collisions at RHIC. The 
    simulation is carried out by binning in centrality according to the number of binary collisions (prescription A: see text for details). Simulations 
    include shadowing and no energy loss. Experimental 
    data are taken from Ref.~\cite{Sahlmueller:2012ru}}
    \label{fig11}
\end{figure}

The first comparisons are carried out for $d$-$Au$ collisions at RHIC energies. 
These experimental results were also historically the first to show the odd effect of an enhancement in peripheral events and a mild suppression in central collisions. 
The data in question are the centrality, $p_T$ and rapidity  (or pseudo-rapidity) dependent nuclear modification factor $R_{dAu}$, defined as, 
\bea
R_{dAu} = \frac{ \int\limits_{b_{min}}^{b_{max}} d^2 b \frac{d^4N_{d Au}}{d^2 p_T d y d^2 b} }
{  \langle N_{bin} (b_{min}, b_{max})\rangle    \frac{d^3 N_{pp}}{ d^2 p_T dy}  } , \label{Raadb}
\eea
where $N$ denotes the yield of leading hadrons or jets, binned in transverse momentum, and rapidity. In the numerator of the above formula, 
one also integrates over a range of impact parameter $b$, which in $d$-$Au$ refers to the 2-D vector from the center of mass  of the large nucleus to the center of mass of the deuteron (in Fig.~\ref{fig3} for example). 
The $\langle N_{bin} (b_{min},b_{max}) \rangle$  in the above formula refers to the mean number of binary nucleon-nucleon collisions per nuclear collision. 

\begin{figure}[htbp]
\resizebox{3.1in}{2.6in}{\includegraphics{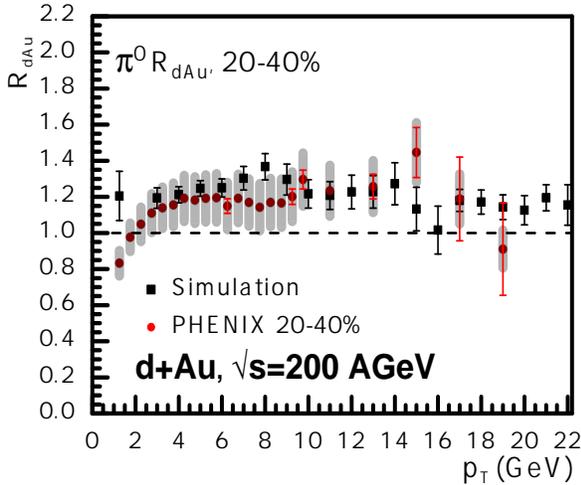}}
    \caption{Color Online: Same as Fig.~\ref{fig11}, except for 20-40\% centrality.}
    \label{fig12}
\end{figure}

As a first step in studying the results of the current simulation in comparison with experimental data, we plot the nuclear modification 
factor for minimum bias collisions. Here no division in centrality bins is carried out and thus there is no discussion of determining 
centrality by number of binary collisions or number of charged particles produced. This serves as a first test of the simulation, which 
performs extremely well in comparison to the data. 
The experimental data have been taken from Ref.~\cite{Sahlmueller:2012ru}. Both the simulation and the experimental data show 
a similar trend: A $p_T$ independent near lack of modification, with the possibility for a minor enhancement between 4 and 16 GeV. 
This is entirely to be expected, high energy jets are mostly unmodified in cold nuclear matter, and the minor enhancement 
can be attributed to the anti-shadowing peak (near $x\simeq 0.1$).  We would further add that in the case of a large centrality 
dependent modification, as is the case in our model (as well as seen in the experimental data), an unmodified minimum bias $R_{dA}$ is by no means 
a trivial outcome.  This is the first hint that the enhancement in peripheral events is being balanced by the suppression in central 
events. 

\begin{figure}[htbp]
\resizebox{3.1in}{2.6in}{\includegraphics{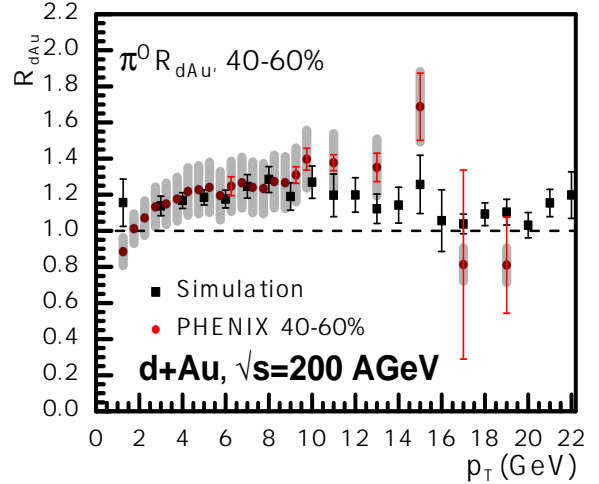}}
    \caption{Color Online: Same as Fig.~\ref{fig11}, except for 40-60\% centrality.}
    \label{fig13}
\end{figure}

The next step is to bin in centrality. Our first attempt will follow convention and utilize the number of binary nucleon-nucleon collisions as 
an indicator of centrality. One runs the nuclear event generator, and 
collects events, classifying them according to the number of binary collisions. One then bins the event according to where $N_{bin}$ lies in 
Fig.~\ref{fig4}. One should point out that while, on average, an increasing $b ( \equiv | \vec{b} |)$ leads to an decrease in $N_{bin}$, any value of $b$ corresponds to a range of binary collisions. This also modifies the numerator of Eq.~\eqref{Raadb}, to
\bea
\mathcal{N}_A \!\!=\!\! \sum_{N_{bin}}\!\! \frac{d^2 N_{d Au}}{d^2 p_T d y } \theta (N_{bin} \!\!- N_{bin}^{min}) \theta(N_{bin}^{max} \!\!- N_{bin}),
\eea
where, $N_{bin}^{min}$ and $N_{bin}^{max}$ are set by the centrality bin that we are interested in. 
The factor of $\langle N_{bin} (b_{min}, b_{max})\rangle $ is simply replaced by $\langle N_{bin} \rangle $ for the 
bin in question, and can be calculated from Fig.~\ref{fig4}. This is referred to as 
prescription $A$ for numerically realizing the numerator of Eq.~\eqref{Raadb}. 

\begin{figure}[htbp]
\resizebox{3.1in}{2.6in}{\includegraphics{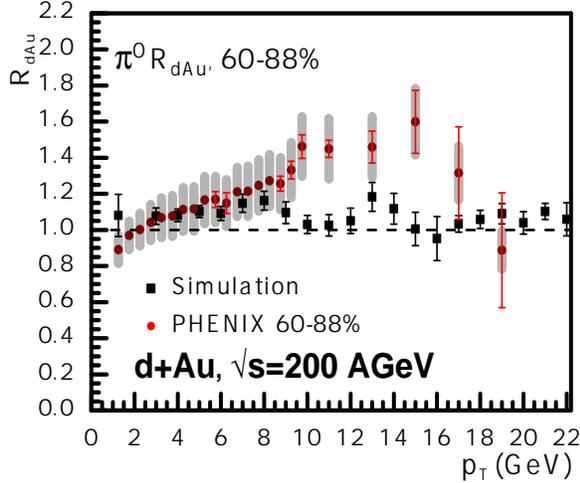}}
    \caption{Color Online:  Same as Fig.~\ref{fig11}, except for 60-88\% centrality. }
    \label{fig14}
\end{figure}

An alternate prescription is to classify events according to the number of produced charged particles, utilizing Fig.~\ref{fig9}
to divide events into different centrality bins. In this case the numerator is replaced with, 
\bea
\mathcal{N}_B \!\!=\!\! \sum_{N_{ch}}\!\! \frac{d^2 N_{d Au}}{d^2 p_T d y } \theta (N_{ch} - N_{ch}^{min}) \theta(N_{ch}^{max} - N_{ch}),
\eea
where, $N_{ch}^{min}$ and $N_{ch}^{max}$ are the minimum and maximum values for charged particles produced, set by the centrality bin that we are interested in. The factor of $\langle N_{bin}\rangle $ in the denominator of Eq.~\eqref{Raadb}, now has to be 
calculated from the collection of events that constitute each centrality bin. We denote this method of calculating the $R_{dA}$ as 
prescription $B$.

\begin{figure}[htbp]
\resizebox{3.1in}{2.6in}{\includegraphics{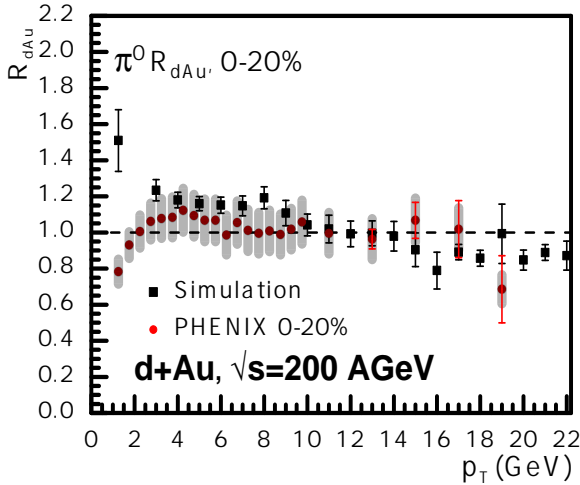}}
    \caption{Color Online:   The nuclear modification factor for neutral pions for 0 - 20\% most central $d$-$Au$ collisions at RHIC. The 
    simulation is carried out by binning in centrality according to the number of charged particles produced (prescription B: see text for details). Simulations 
    include shadowing and no energy loss. Experimental 
    data are taken from Ref.~\cite{Sahlmueller:2012ru}.}
    \label{fig15}
\end{figure}

As most readers are aware, prescription $A$, is the usual theoretical method of calculating the centrality dependence of the nuclear modification factor, whereas prescription $B$ is closer to the experimental method of determining centrality. We first show the results of simulating the centrality dependence of the 
pion $R_{dA}$ using prescription $A$ or using the number of binary collisions.

\begin{figure}[htbp]
\resizebox{3.1in}{2.6in}{\includegraphics{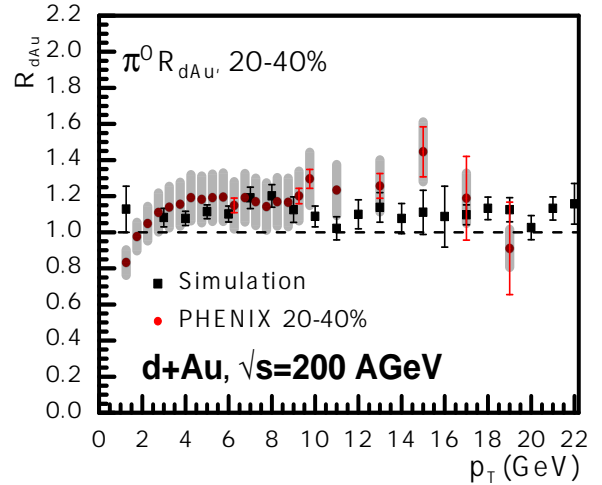}}
    \caption{Color Online:  Same as Fig.~\ref{fig15}, except for 20-40\% centrality. }
    \label{fig16}
\end{figure}

In Fig.~\ref{fig11}, the $R_{dA}$  for the top 0-20\% most central collisions are plotted. One immediately notes an enhancement in the
simulation, but no such enhancement in the experimental data, which seem to be consistent with unity. The simulation does not explain the experimental 
data. The enhancement in central events such as demonstrated by the simulation, is entirely expected based on the shadowing function that has been 
used to generate events. Within this framework, the complete lack of any modification in the experimental data is rather surprising; central event should 
present the maximal nuclear modification.

\begin{figure}[htbp]
\resizebox{3.1in}{2.6in}{\includegraphics{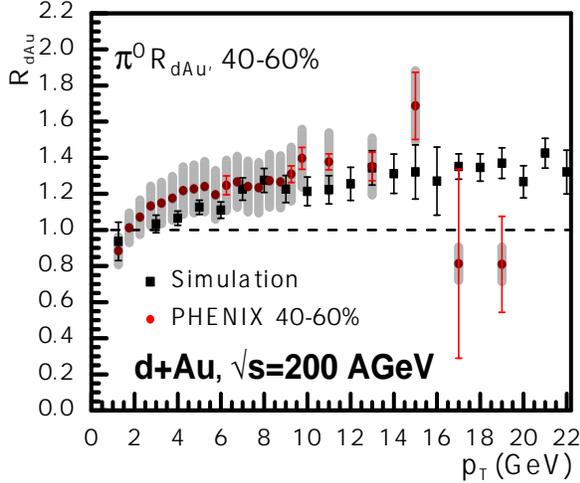}}
    \caption{Color Online:  Same as Fig.~\ref{fig15}, except for 40-60\% centrality. }
    \label{fig17}
\end{figure}

As one moves up in centrality, from most central to peripheral events, the enhancement seen in the simulation tends to reduce progressively. 
There is less enhancement in the 20-40\% events, even less in the 40-60\% simulations, with no modification at all in the 60-88\% events. 
This behavior of the simulation is entirely expected, as we move from cases with the largest expected nuclear density modification to cases with little density and 
hence no modification at all in the $R_{dA}$. The experimental data, however, show an entirely different trend: with no modification in the central 
event and the $R_{dA}$  rising with centrality from most central to most peripheral events. The fact that the simulation results with prescription $A$ 
match some of those from the experiment is entirely coincidental. The simulation for the $R_{dA}$ drops as one transitions from central to peripheral while the 
data trend in the opposite direction.

\begin{figure}[htbp]
\resizebox{3.1in}{2.6in}{\includegraphics{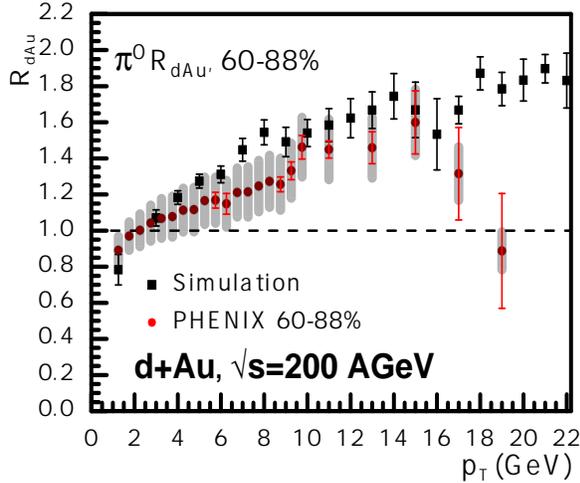}}
    \caption{Color Online:  Same as Fig.~\ref{fig15}, except for 60-88\% centrality. }
    \label{fig18}
\end{figure}

The experimental results for $R_{dA}$ in $d$-$Au$ collisions are rather unexpected. The largest modification is seen in the most peripheral bin, which 
by all accounts should resemble $p$-$p$ most closely. We now attempt to calculate the $R_{dA}$ using prescription $B$, i.e., using the simulated number of charged 
particles produced to bin in centrality. The charged particles are gathered over all rapidities, in events that contain a high-$p_T$ $\pi^0$ and then 
compared with the outlined division in Fig.~\ref{fig9}. Using this prescription, an excellent agreement is obtained with experimental data on the nuclear 
modification factor of high $p_T$ neutral pion production. One notes that for central collisions, the $R_{dA}$ is consistent with one and continues to rise 
as one moves towards more peripheral collisions.

\begin{figure}[htbp]
\resizebox{3.4in}{3in}{\includegraphics{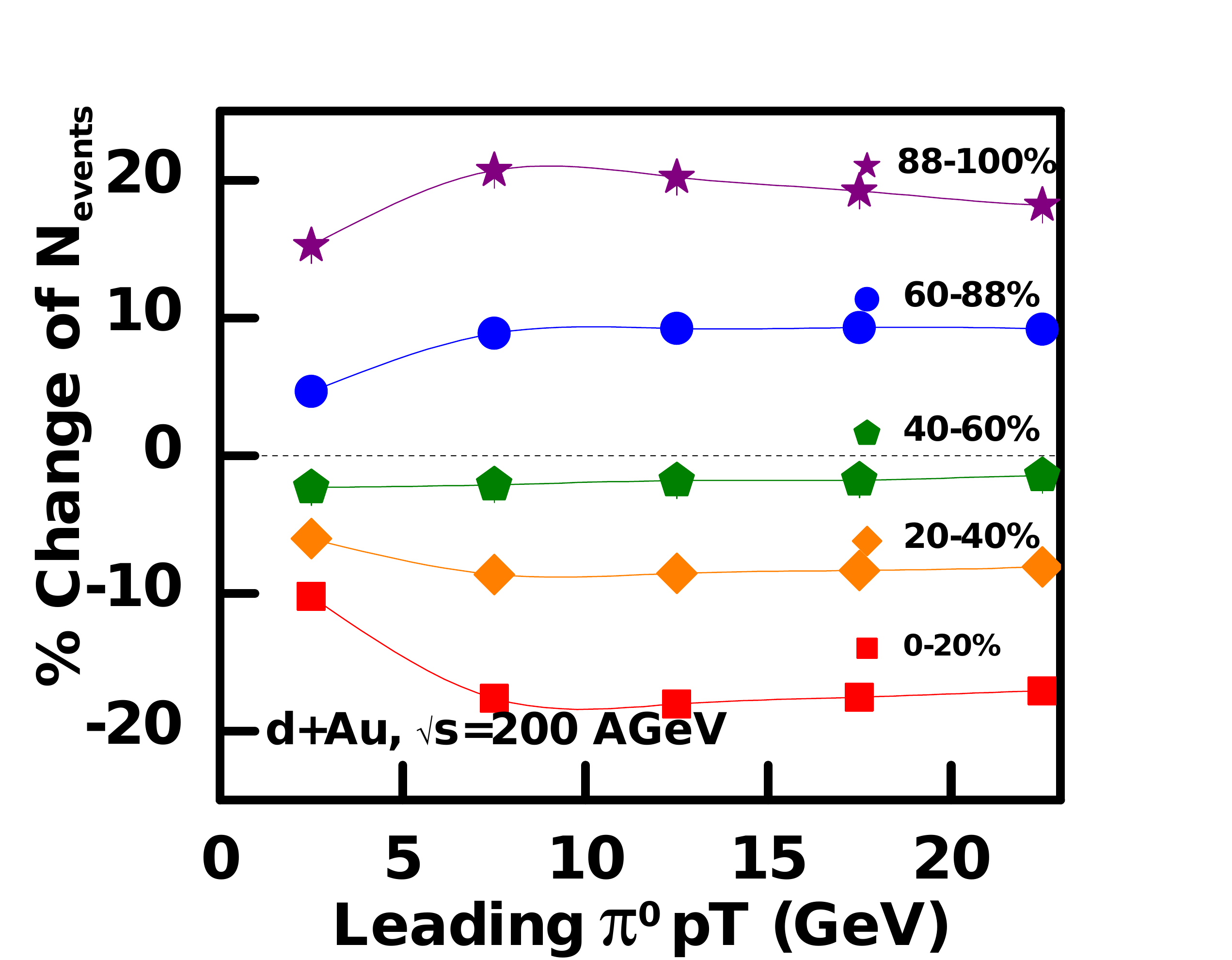}}
    \caption{Color Online: The fraction of events that shift in or out from each centrality bin as the definition of centrality is changed from binary collisions to number of charged particles produced. The fractional bin shift is plotted as a percentage of the number of events in the original definition with number of binary collisions, as a function of the transverse energy of the detected pion. See text for details.}
    \label{fig19}
\end{figure}

To understand the reason behind the positive comparison between simulation and experiment, we focus on how the events with jets are binned in 
different centrality bins. In particular we look at how the number of events within each bin, change as we transition from binning according to the number of 
binary collisions to binning according to the number of charged particles produced. We focus on events with a high $p_T$ pion and isolate the number of events captured in each centrality
bin defined by the number of charged particles produced (prescription $B$), subtracted from this is the number of events captured in the same bin defined by the number of binary collisions (prescription $A$). This difference is then expressed as a fraction of the number of events captured using prescription $A$. 
This is plotted as a function of the $p_T$ of the pion in Fig.~\ref{fig19}. We notice that central and the number of  semi-central (20-40\%) events when binned in terms of produced charged particles are
suppressed compared to the case when they binned according to the number of binary collision. These lost events show up in the more peripheral collisions, and 
lead to an enhancement in those collisions. This is the reason that peripheral events as measured in experiment are enhanced compared to binary scaled $p$-$p$. 
Central collisions, compared to binary scaled $p$-$p$ are slightly enhanced due to shadowing. These lose events to peripheral collisions and as such the yield is 
reduced, leading to the ratio of central collisions to binary scale $p$-$p$ to be close to unity. 

\begin{figure}[htbp]
\resizebox{3.4in}{3in}{\includegraphics{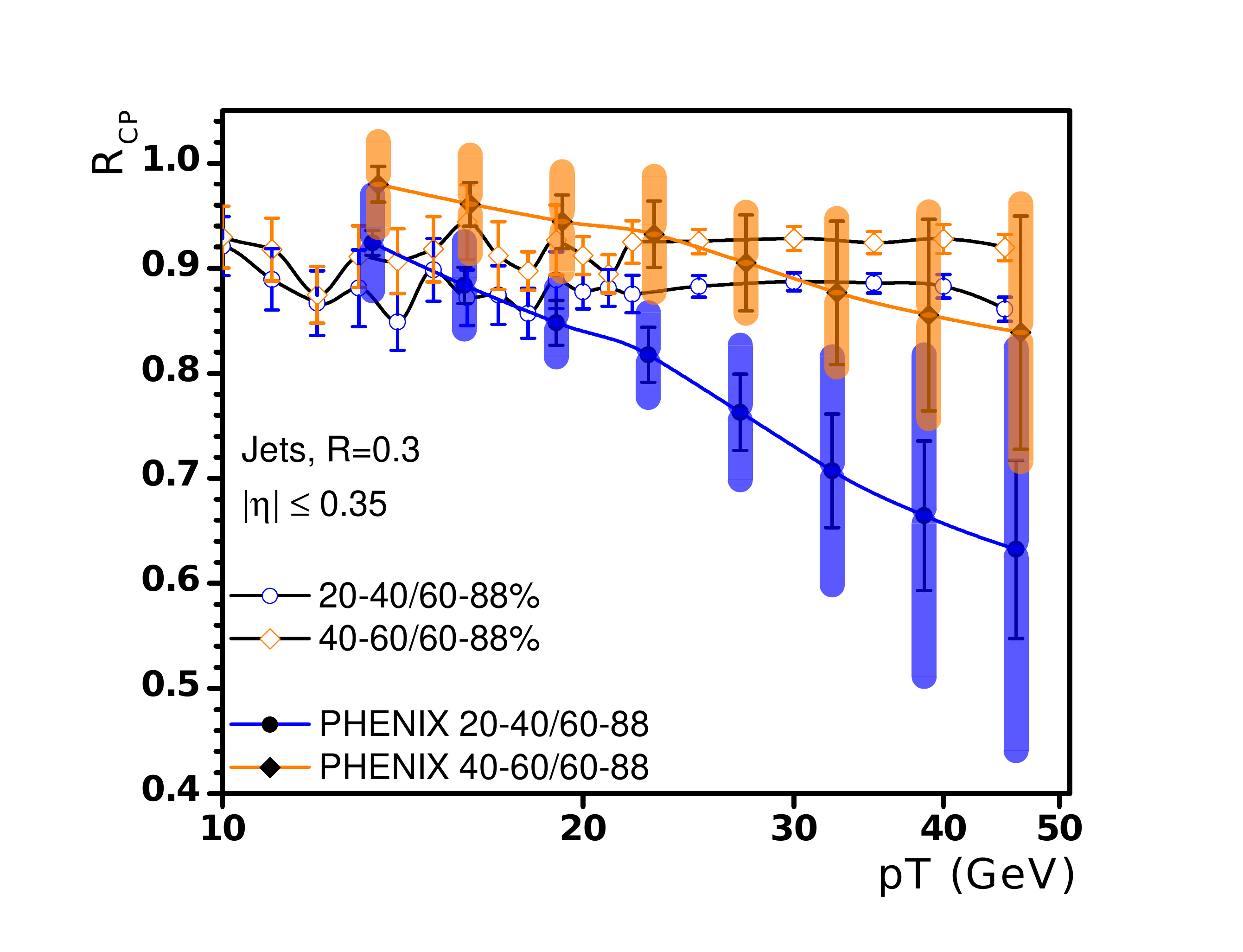}}
    \caption{Color Online: The ratio of the nuclear modification factor of jets produced in $d$-$Au$ collisions at RHIC.
 Experimental data are taken from Ref.~\cite{Adare:2015gla}.}
    \label{fig20}
\end{figure}

This ``movement'' of events from central to less central to peripheral collisions, leads to an enhancement over the expected yield in more peripheral collisions, and 
a suppression over the expected enhancement in central events. This is mostly an initial state effect. In events with a high $p_T$ $\pi^0$, there has to be a high-$x$
parton in the initial state of at least one nucleon in both the $d$ and the $Au$ nuclei. The presence of a large-$x$ parton in a nucleon of the $d$ depletes the amount of energy available to produce several additional soft partons and as such the collisions of this nucleon with nucleons in the $Au$ leads to the production of fewer charged hadrons. This in turn leads to this event being binned as a more peripheral event. To test this concept further, we plot the jet $R_{CP}$, the ratio of the jet spectrum in central to peripheral events, both scaled by the number of expected binary collisions, as a function of 
$p_T$ in Fig.~\ref{fig20}.  The results of these simulations are close to experiment if the error bars are taken into consideration.  There is some concern with the 0-20\% central data as it does not appear to be consistent between jet and pion measurements, therefore it was omitted from the plot.  In addition, the differences in the methods to determine centrality as well as in reproducing jets between this simulation and experiment could account for the observed separation.
The $R_{CP}$ is suppressed compared to unity as events move out of more central bins towards more peripheral events. This same effect is transferred via fragmentation to the $\pi^0$ and manifests in the $R_{dA}$ as discussed earlier. 

\begin{figure}[htbp]
\resizebox{3.4in}{3in}{\includegraphics{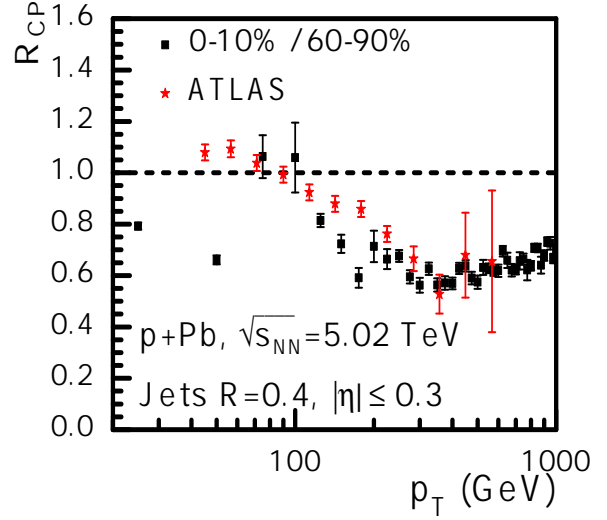}}
    \caption{Color Online: The ratio of the nuclear modification factor of jets produced in $p$-$Pb$ collisions at the LHC.
 Experimental data are taken from Ref.~\cite{ATLAS:2014cpa}.}
    \label{fig21}
\end{figure}

The primary question at this point is if this effect is solely driven by energy conservation: Is the reduced energy available for the production of small-$x$
partons the only reason for the reduction in the charged particle production, or is there a multi-particle correlation which leads to fluctuation with fewer hard partons, versus fluctuations to several soft partons. In the standard language of pQCD these would be considered as higher-twist multi-parton distribution functions. 
In an alternative formalism, we ask if this is being caused due to an initial state \emph{color transparency}~\cite{Brodsky:1994kf,Frankfurt:1993it,Frankfurt:1988nt}: The fluctuation of the nucleon to a smaller state with fewer hard partons. In order to study this question further, we consider the modification of this process with energy of the collision and with the energy of the jet. The higher the energy 
of the jet, the larger the $Q^2$ of the process, and as a result, the smaller the size of the fluctuations will be in the proton. This should lead to a more pronounced effect in similar observables at LHC energies with jets or leading hadrons at much higher energies. 

\begin{figure}[htbp]
\resizebox{3.4in}{3in}{\includegraphics{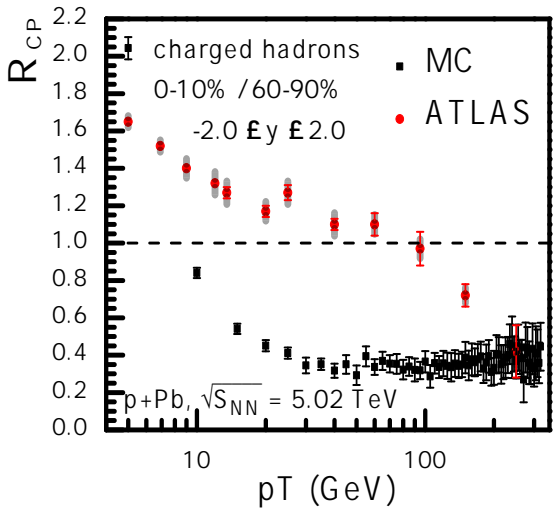}}
    \caption{Color Online: The ratio of the nuclear modification factor of charged particles produced in $p$-$Pb$ collisions at the LHC.
 Experimental data are taken from Ref.~\cite{ATLAS:2014cza}.}
    \label{fig22}
\end{figure}

In Fig.~\ref{fig21}, we plot the $R_{CP}$ of jets in central $p$-$Pb$ collisions at mid-rapidity, measured by the ATLAS collaboration at the LHC. 
This represents the ratio of the nuclear modification factor in central events (0-10\%) to that in peripheral events (60-90\%).
At high energies, where the effect of energy conservation should become important, our simulation, once again compares very well with the data.
In Fig.~\ref{fig22}, we plot the $R_{CP}$ for charged particles in a similar range of centrality between central and peripheral collisions. In this case, 
while we obtain the magnitude of the suppression, we do not get the shape of the experimental $R_{CP}$ data. Given that on the jet side, the 
agreement between simulation and experiment starts around $60$GeV, the disagreement between the simulation and data for the $R_{CP}$ of 
charged particles below a 100GeV is somewhat puzzling. It should be pointed out that in these cases, we are dealing with a larger number of partons
produced, all of which are color correlated. The effect of this on the fragmentation of the leading parton has not been studied in this effort. 
As a result,  on the basis of these results, we cannot definitely state whether or not color 
transparency plays a role in these measurements. 

At the risk of repetition, we point out again that our simulations do not in any way contain rescattering and secondary particle production. In 
the interest of keeping the hard particle production as close to reality as possible (without the need for artificial shadowing), 
we have abandoned the energy enhanced PDF for the partons
in the struck nucleus. There are thus many points of departure between our simulations and the experimental data on soft particle production. 
Our goal in this effort was to point out that events with a hard jet have a lower soft particle production rate, which leads to binning in a more peripheral 
bin. While this goal is now firmly established, this work should, by no means, be considered definitive, as our efforts to determine whether 
color transparency plays a role in these collisions, beyond energy conservation, has not yielded a clear response. These and other topics will be 
discussed at length in the subsequent section.

\section{Discussion and Outlook} \label{outlook}

In this paper, new experimental results from both RHIC and LHC on jet production in extremely asymmetric systems has been discussed. 
At both the energy scales of RHIC and LHC, similar results were discovered: events that contained a jet or a high energy particle, seemed to show an enhancement over binary scaled $p$-$p$ in peripheral events and a suppression compared to the expectation of shadowing and 
binary scaling enhanced central collisions. Our goals in this effort were two fold: The first goal was to set up a reliable event generator that could be used to reproduce some portion of the observed experimental data from such collisions. Based on the success of this event generator, our second goal was to determine if 
the observed behavior can solely be explained by energy conservation or if it requires the incorporation of correlations similar to that of color transparency. 

The designed parameter free event generator consisted of two parts: A nuclear Monte-Carlo to determine the positions of the nucleons within the 
nucleus, and a modified version of PYTHIA, with an event-by-event shadowing and PDF enhancement to account for the collision of 
a nucleon from the $p(d)$ with a column of nucleons within the larger nucleus. 
The results from these simulations, manage to correctly predict the behavior in both the 
jet $R_{CP}$ and leading particle $R_{dA}$ at RHIC, and the jet $R_{CP}$ at the LHC. The 
simulation also correctly predicts the magnitude of the suppression in the leading particle $R_{CP}$ at the LHC, though it does not reproduce the 
shape of the curve. This is a considerable success for such an endeavor. 
The event generator presented in this effort, cannot be considered as complete, there remain several soft observables that, with the given 
setup of not containing an energy enhanced PDF and without rescattering corrections, cannot be explained. 
In spite of these, the above study will greatly inform the design of future event generators which will have to be set up to explain these 
striking experimental data. While this simulation was built on top of the $p$-$p$ generator PYTHIA, future generators that incorporate 
all of the above insights will have to be built as a more original effort.

Our goal of setting up the current generator  (as well as future generators) 
was to use it to extract the physics underlying these new observations. These simulations have now 
established the notion that the enhancement in peripheral events and suppression in central events is entirely due to suppression in soft 
parton production in a nucleon with a large-$x$ parton. A large portion of this is entirely due to the reduced energy available for soft 
parton production. Is there any further correlation due to color transparency like effects? The fact that our $Q^2$ independent shadowing 
led to a successful description of the jet $R_{CP}$ at the LHC would seem to rule out such an effect. However, the simulation did not manage 
to explain the shape of the leading particle $R_{CP}$. Note that both the leading particle $R_{dA}$ and the jet $R_{CP}$  at RHIC energies were mostly accounted for by the simulation.  In order to study such a correlation in greater detail, one needs to devise an event generator which will incorporate an energy enhanced PDF, with a far more sophisticated shadowing set up to reproduce the large-$x$ behavior of the PDF within a single nucleon. 
We leave the set up of such an event generator for a future effort. Alternatively, a mechanism will have to be set up where the PDF of the nucleon (or nucleons) 
from the projectile will have to be sampled once in a $p(d)$-$A$ collision.

Beyond the study of such initial state color transparency effects, a future more advanced event generator for asymmetric collisions such as $d$-$Au$ 
or $He^3$-$Au$ will also allow for a deeper understanding of the quantum correlation between nucleons in a nucleus. In the current work, we have explored 
excluded volume corrections in a  Woods-Saxon distribution, as well as Gaussian perturbations in a shell model based distribution. Experimental 
data, coupled with theory uncertainties at the partonic level do not allow to distinguish between the different correlations between nucleons. 
However, these can be studied systematically, once the partonic component is settled via $p$-$A$ collisions. This will allow an extension of 
nuclear structure which has so far not been extensively studied.  

Extremely asymmetric nuclear collisions with a hard interaction provide a new window into a large variety of correlation phenomena at multiple 
scales. Future studies with more accurate experimental data, as well as a more sophisticated event generator, will reveal new information regarding the 
correlation between partons within a single nucleon, as well as correlations between nucleons in large nuclei. The current work represents a bench mark in this 
direction, providing a glimpse of the insights that may be gained by such a research program as well as highlighting the ingredients and framework required for future efforts.

\acknowledgements
 The authors would like to thank S. Gavin, G.-Y. Qin and members of the JET collaboration for helpful discussions.
A.M. is indebted to G. David and M. A. Lisa, for discussions that lead to the genesis of this work. 
This work was supported in part by the National Science Foundation under grant number PHY-1207918, and by the Office of Science of the US Department of 
Energy under grant number DE-SC0013460.

\bibliography{refs}

\end{document}